\documentclass[twocolumn]{aastex631}
\usepackage{graphicx}
\usepackage{xcolor}
\usepackage [english]{babel}
\usepackage [autostyle, english = american]{csquotes}
\usepackage{amsmath}
\usepackage{makecell}
\usepackage{booktabs}
\usepackage{multirow}
\MakeOuterQuote{"}

\begin{document}

\title{AGN Feedback Models and AGN Demographics II: Comparing Predictions of Radiative and Total Feedback to Observations}

\author{Arjun Suresh}
\author{Michael R. Blanton}
\affil{New York University}

\begin{abstract}

We evaluate the radiative--mode active galactic nucleus (AGN) feedback models of EAGLE, SIMBA, and TNG100 by comparing their predictions for the AGN--host galaxy relationship to new observational constraints. Owing to incomplete knowledge of the underlying physics, these models differ substantially, and it remains unclear whether any of them accurately reflect reality. In a previous study, based on the demographics of narrow line AGN, we constrained $F_{\rm AGN}$, the intrinsic fraction of galaxies hosting a radiative AGN with Eddington ratio $\lambda > 10^{-3}$, as a function of host stellar mass ($M_\star$) and specific star formation rate (sSFR). Observationally, $F_{\rm AGN}$ declines strongly with $M_\star$ for quiescent galaxies, while remaining approximately constant for star-forming systems. In this study, we find that none of the simulations reproduce these trends even qualitatively, indicating a mismatch between the simulated and observed radiative AGN populations. Additionally, since EAGLE does not explicitly distinguish between radio and radiative modes, we compare its predictions for $F_{\rm AGN}(M_\star, \mathrm{sSFR})$ to our novel measurements of the combined radiative and radio mode AGN fractions. For EAGLE’s constant coupling efficiency $\epsilon_{\rm f} = 0.15$, and for the adopted standard bolometric and jet-power conversion factors in the observations, the total AGN feedback energy in EAGLE is broadly consistent with observations, though with Eddington ratios a factor of 10 greater than is observed. More detailed comparisons of EAGLE’s assumptions with observations are therefore required before drawing firm conclusions.

\end{abstract}

\section{Introduction}\label{intro}

Nearly all massive galaxies are thought to host a supermassive black hole at their center (\citealt{magorrian1998}), though only some display detectable emission. These objects are called active galactic nuclei (AGN) and can be highly luminous across the electromagnetic spectrum. On the basis of both theoretical and observational studies of X-ray binaries and AGN, the consensus in the literature is that AGN operate in two distinct major modes of activity --- radiatively efficient and radiatively inefficient (\citealt{maccarone2003,merloni08,heckman14a,MHD03}). The radiatively efficient mode is believed to occur at high accretion rates ($L_{\rm Bol}/ L_{\rm Edd} > 0.02$) and the radiatively inefficient mode at low accretion rates ($L_{\rm Bol}/ L_{\rm Edd} < 0.02$). Here, $L_{\rm Bol}$ is the bolometric luminosity of the AGN and $L_{\rm Edd}$ is the Eddington luminosity of the black hole. In this work, we focus on radiatively efficient AGN, comparing their observed relationships with host galaxies, to predictions of the same from simulations, as a means of testing the black hole subgrid models they employ. This study complements \cite{SureshBlanton2026}, which examined the radiatively inefficient AGN population.

Cosmological simulations commonly invoke active galactic nucleus (AGN) feedback as a key component of galaxy formation models in order to reproduce the observed quenched population of galaxies (\citealt{somerville&davee2015, Dimatteo2005, silk&rees1998}). In these models, AGN feedback acts to suppress star formation in massive galaxies by injecting energy and momentum into the host galaxy’s interstellar medium. Radiatively efficient AGN---often described as operating in the ``radiative'' or ``quasar'' mode---are thought to influence their host galaxies primarily through intense radiation fields and radiatively driven winds. These processes can expel gas from the galaxy, thereby depleting the reservoir of cold gas available for star formation. In contrast, radiatively inefficient AGN---commonly referred to as operating in the ``radio'' mode---are believed to suppress star formation by inhibiting the cooling of hot gas, thereby preventing the replenishment of the cold gas supply needed to sustain star formation.

Although there isn't yet direct observational evidence that AGN feedback plays the dominant role in regulating galaxy formation, the framework is still strongly supported by a broad range of observational studies (\citealt{Harrison&Almeida2024, fabian12}). Numerous studies have reported the presence of galactic-scale outflows spanning multiple gas phases. These include hot, ionised ultra-fast outflows \citep{Tombesi2013, chartas2021}, warm ionised outflows \citep{Harrison2014, foster2014}, neutral atomic outflows \citep{morganti2016, maiolino2019}, and cold molecular outflow components \citep{Rupke_2013, González-Alfonso_2017, cicone2014, fluetsch2018}. However, most observational studies only focus on a single outflow phase, with key properties such as mass, velocity, and energy often subject to large uncertainties. As a result, a unified picture of AGN feedback remains elusive, partly due to small, inhomogeneous samples that are biased toward the most extreme AGN systems.

From the relatively small number of studies that have characterised multi-phase outflows in AGN samples (e.g. \citealt{riffel2023, Rupke_2017}), cold molecular gas is found to dominate the total outflow mass, followed by the neutral atomic phase and, subsequently, the warm ionised component. In the local Universe, molecular outflows are typically slower and more compact than the ionised outflows, with velocities of $v \lesssim 300$--$500~\mathrm{km\,s^{-1}}$ and spatial extents of $r \lesssim 1$--$2~\mathrm{kpc}$. In contrast, ionised outflows generally reach higher velocities of $\sim 500$--$1000~\mathrm{km\,s^{-1}}$ and extend over larger spatial scales, with radii of $r \sim 1$--$10~\mathrm{kpc}$ (\citealt{Harrison&Almeida2024}).

Assessing the global and long-term impact of AGN activity on star formation through statistical studies of AGN and galaxy samples is challenging, owing to the mismatch between the characteristic timescales of the two processes. AGN activity can vary substantially on very short timescales (down to days), whereas star formation rates typically trace much longer timescales of order $\sim 100~\mathrm{Myr}$. Nevertheless, there exists observational evidence that AGN-driven outflows can locally influence the star formation rates (SFRs) of their host galaxies on shorter timescales. Using IFU and ALMA observations, \cite{Bessiere&Almeida2022} report spatial correlations between AGN-driven outflows and both enhancements and declines in flux associated with young stellar populations. Additionally, several studies have found lower nuclear star formation rates or reduced nuclear molecular gas fractions in AGN hosts compared to matched control samples of inactive galaxies \cite{Ellison2021, Lammers2023, Rosario2019, Garcia2021, Garcia2019}.

Taken together, this body of observational evidence, along with the fact that cosmological simulations can reproduce the observed galaxy population only when AGN activity is incorporated, has established AGN feedback as a fundamental ingredient of modern galaxy formation models and the leading theoretical explanation for the quenching of star formation in massive galaxies. Furthermore, beyond galaxy quenching, AGN feedback is also believed to play a critical role in several other phenomena that are central to our understanding of the Universe. For instance, by expelling gas from galaxies to group and cluster scales, AGN feedback is thought to substantially suppress the nonlinear matter power spectrum (\citealt{vandaalen2011}). Accurately characterizing and constraining this effect is essential for weak-lensing tests of $\Lambda \mathrm{CDM}$ cosmology and for distinguishing it from signatures of beyond-standard-model physics that can produce similar observational effects (\citealt{bigwood2025}). Finally, a robust understanding of AGN activity is crucial for explaining the growth and, ultimately, the origin of supermassive black holes.

Clearly, an accurate characterization of AGN feedback is indispensable for understanding a wide range of astrophysical and cosmological processes. Nevertheless, as noted earlier, existing observational constraints remain insufficient to fully elucidate the physical mechanisms that drive AGN feedback. As a result, the sub-grid prescriptions employed in cosmological simulations, although motivated by physical considerations, remain largely phenomenological and are often calibrated to reproduce selected observed galaxy properties. Given the far-reaching implications of these models for our understanding of the Universe, conducting independent observational tests is essential in order to validate whether the physics they employ is a faithful reflection of reality. Despite their importance, independent observational tests of AGN sub-grid models remain scarce, with only a handful of examples in the literature (see \citealt{houda2022, habouzit2022}). But even in those studies, the observational cosntraints they use have not all been corrected for the relevant selection effects.

To remedy the above situation, starting with \cite{SureshBlanton2024} we devised a set of studies to make accurate measurements of the AGN-host relationship. The whole program relies heavily on optical integral field unit data from the Mapping Nearby Galaxies at Apache Point Observatory (MaNGA; \citealt{bundy15a}), a component of the Sloan Digital Sky Survey IV (SDSS-IV; \citealt{blanton17a}). The high-precision measurements provided by MaNGA allow for a rigorous characterization of the flux and star formation rate-related selection effects that have historically hindered accurate assessment of the intrinsic AGN–host galaxy connection that exists in the universe. The AGN-host relationships are quantified in terms of model fits to the Eddington ratio distribution of AGN, as a function of host galaxy $M_\star$ and sSFR. We then evaluate the model fits to get $F_{\rm AGN}$ --- the completeness--corrected estimate of the fraction of galaxies that host an AGN with an Eddington ratio $\lambda > \lambda_{\rm c}$, as a function of host galaxy $M_{\star}$ and sSFR. $F_{\rm AGN} \left(M_{\star}, sSFR \right)$ is a simple yet robust independent observational constraint to test predictions by simulations against. In \cite{SureshBlanton2024}, we measured $F_{\rm AGN}(M_{\star}, {\rm sSFR})$ for the local radio AGN population. We then used these measurements in \citet{SureshBlanton2026} to test the predictions of radio-mode AGN feedback in cosmological simulations.

In \cite{blanton2026}, we extended this analysis to the local radiative AGN population by measuring $F_{\rm AGN}(M_{\star}, {\rm sSFR})$ for narrow-line AGN with $\lambda > \lambda_{\rm c} = 10^{-3}$. Here, the Eddington ratio is defined as $\lambda = L_{\rm Bol}/L_{\rm Edd}$, where $L_{\rm Bol}$ is the bolometric luminosity of the AGN and $L_{\rm Edd}$ is the Eddington luminosity of the black hole. In that study, we found that:

\begin{itemize}
    \item For star forming galaxies ($ \log \left( sSFR/yr^{-1} \right) > -11.5$),  $\log_{\rm 10} \left(F_{\rm AGN} \right)$ is fairly constant with stellar mass across the range $10 \leq \log_{\rm 10} \left( M_\star / M_\odot \right) \leq 12$ 

    \item For quiescent galaxies ($ \log \left( sSFR/yr^{-1} \right) < -11.5$), $\log_{\rm 10} \left(F_{\rm AGN} \right)$ decreases strongly with stellar mass across the range $10 \leq \log_{\rm 10} \left( M_\star / M_\odot \right) \leq 12$ 
\end{itemize}

\begin{figure*}[h]
    \centering
    \includegraphics[width = 7.2in]{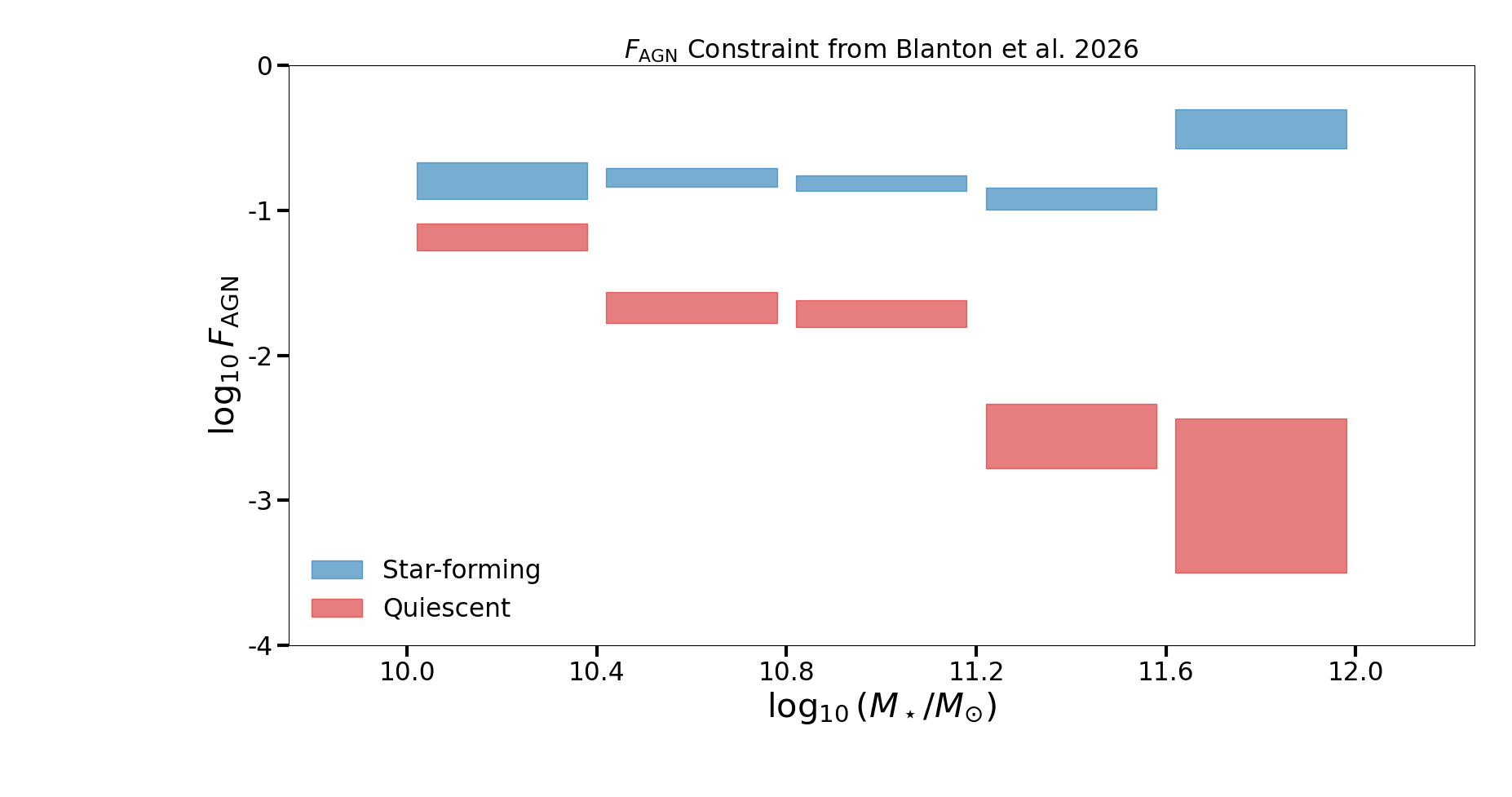}
    \caption{Observed intrinsic AGN fraction ($F_{\rm AGN}$) as a function of $M_{\star}$ and sSFR. Here, $F_{\rm AGN}$ represents the fraction of galaxies hosting a narrow-line AGN with an Eddington ratio $\lambda > 10^{-3}$. Data are derived from the MaNGA optical IFU catalog, adapted from Appendix A of \cite{blanton2026}. In this paper, we test predictions from the EAGLE, SIMBA, and TNG100 against these observed $F_{\mathrm{AGN}}$ trends, as a means of evaluation of their underlying subgrid physics.}
    \label{fig:Fagn_constraint}
\end{figure*}

\begin{table*}[t]
    \centering
    \begin{tabular}{ccccc}
        \toprule
        \multirow{2}{*}{\textbf{$\log_{\rm 10} \left( M_\star/ M_\odot\right)$}} & \multicolumn{2}{c}{$\log F_{\text{AGN}}^{\text{SF}}$} & \multicolumn{2}{c}{$\log F_{\text{AGN}}^{\text{Q}}$} \\ 
        \cmidrule(lr){2-3} \cmidrule(lr){4-5}
                                           & Lower Bound & Upper Bound & Lower Bound & Upper Bound \\ 
        \midrule
        10 -- 10.4        & $-0.924$     & $-0.67$     & $-1.28$     & $-1.09$     \\ 
        10.4 -- 10.8      & $-0.84$     & $-0.71$     & $-1.78$     & $-1.56$     \\ 
        10.8 -- 11.2      & $-0.87$     & $-0.76$     & $-1.81$     & $-1.62$     \\ 
        11.2 -- 11.6      & $-1.00$     & $-0.84$     & $-2.78$     & $-2.34$     \\ 
        11.6 -- 12        & $-0.57$     & $-0.30$     & $-3.50$     & $-2.44$     \\ 
        \bottomrule
    \end{tabular}
    \caption{1$\sigma$ bounds of $\log_{\rm 10} \left(F_{\rm AGN} \right)$ in bins of $\log_{\rm 10} \left( M_\star/ M_\odot\right)$
    for the radiative AGN observational constraint, for star forming and quiescent galaxies.}
    \label{tab:radiative_Fagn_constraint}
\end{table*}

Figure~1, reproduced from Appendix~A of \cite{blanton2026}, shows the above results and Table \ref{tab:radiative_Fagn_constraint} provides the values. We note that \cite{aird2019} and \cite{birchall2023} made similar measurements of the radiative AGN fraction using X-ray emission rather than narrow emission lines. Their results for the $F_{\rm AGN}$ trends in star-forming galaxies qualitatively agree with ours, although there are substantial quantitative differences. However, their results for quiescent galaxies qualitatively disagree with ours, as they do not find a decline in $F_{\rm AGN}$ with $M_{\star}$. The origin of these differences remains unclear, and direct comparisons between the two studies are necessary. See \citet{blanton2026} for further discussion.

Despite these discrepancies, we proceed using our narrow-line measurements as the observational benchmark in this study. Cosmological simulations, if they truly employ AGN sub-grid models that faithfully reflect reality, should at least qualitatively reproduce the observed trends in $F_{\rm AGN}(M_{\star}, {\rm sSFR})$. Here, we compare predictions for $F_{\rm AGN}(M_{\star}, {\rm sSFR})$ from the EAGLE (\citealt{crain2015, schaye2015}), SIMBA (\citealt{simba}), and TNG100 (\citealt{pillepich18, illustristng2019}) cosmological simulations with our observational measurements of the same quantity. The black hole feedback models in these three simulations have been calibrated to reproduce the following observables:

 \begin{itemize}
     \item EAGLE: the observed galaxy stellar mass function at z = 0.1 (\citealt{li&white2009, baldry12}) and the observed galaxy size-mass relationship of spirals at z = 0.1 (\citealt{shen2003, baldry12})
     \item SIMBA: the observed galaxy stellar mass function at z = 0 (\citealt{bernardi17}) and the observed $M_{\rm BH}-M_{\star}$ scaling relation (\citealt{K&H2013, bentz2018})
     \item TNG100: the observed galaxy stellar mass function at z = 0 (\citealt{baldry2008, bernardi2013}), the global star formation rate density as a function of cosmic time (\citealt{behroozi13, oesch2015}) and the $M_{\star}$-$M_{\rm halo}$ relationship at z = 0 (\citealt{behroozi13, moster2013})
 \end{itemize}

Note that none of the above simulations have been tuned to reproduce the observed AGN-host relationship in terms of $F_{\rm AGN} \left(M_{\star}, ~ \rm sSFR \right)$. Consequently, this study serves as an independent test of the black hole feedback models employed in these simulations. 

In Section \ref{sec:sims_and_bhfeedback} we briefly describe the EAGLE, SIMBA and TNG100 cosmological simulations and summarize their black hole feedback models. In Section \ref{sec:data_and_methods} we describe the datasets and we use and methods of analysis we employ in this study. In Section \ref{sec:results} we report our findings of the independent tests of the AGN feedback models of EAGLE, SIMBA and TNG100 using our observed $F_{\rm AGN} \left( M_{\star}, sSFR\right)$ constraint. Finally, in Section \ref{sec:conclusion} we summarize our conclusions.

\section{Cosmological Simulations and their Black Hole Feedback Models}\label{sec:sims_and_bhfeedback}
\subsection{Cosmological Simulations}\label{sims}
We previously summarized the EAGLE, SIMBA, and TNG100 cosmological hydrodynamical simulations in \cite{SureshBlanton2026}, so we provide only a brief overview here. For comprehensive descriptions of the simulations, we direct the reader to \cite{schaye2015} and \cite{crain2015} for EAGLE, \cite{simba} for SIMBA, and \cite{pillepich18} and \cite{illustristng2019} for TNG100. In this study, we utilize the primary flagship run of each simulation suite i.e., Ref-L0100N1504 for EAGLE, the $100\ \rm cMpc$ flagship run for SIMBA, and TNG100-1 for TNG. 

All three simulations assume a standard $\Lambda$ cold dark matter ($\Lambda$CDM) cosmology with parameters broadly consistent with Planck constraints: $\Omega_{\Lambda} \approx 0.70$, $\Omega_{\rm m} \approx 0.30$, $\Omega_{\rm b} \approx 0.05$, and $h \approx 0.68$ (\citealt{planck_collab2014}). Furthermore, each model tracks the coupled evolution of dark matter, gas, stars, and supermassive black holes. Due to limited resolution, all three simulations employ sub-grid recipes for star formation, stellar feedback, black hole accretion, black hole feedback, etc. The salient features and parameters of the simulation runs used in this work are summarized in Table~\ref{tab:sims_summary}.

\begin{table*}[t]
    \centering
    \caption{Summary of key parameters and features across the EAGLE, SIMBA and TNG100 simulations. Columns denote the simulation and run name, the base hydrodynamics solver, the comoving box length ($L_{\rm box}$), the total number of initial particles/cells ($N_{\rm part}$), gas mass resolution ($m_{\rm gas}$), dark matter mass resolution ($m_{\rm DM}$), and the starting ($z_{\rm max}$) and ending ($z_{\rm min}$) redshifts.}
    \label{tab:sims_summary}
    \begin{tabular}{llcccccc}
        \toprule
        Simulation (Run) & Hydro Code & $L_{\rm box}$ & $N_{\rm part}$ & $m_{\rm gas}$ & $m_{\rm DM}$ & $z_{\rm max}$ & $z_{\rm min}$ \\
        & & ($\rm cMpc$) & & ($\rm M_{\odot}$) & ($\rm M_{\odot}$) & & \\
        \midrule
        EAGLE (Ref-L0100N1504) & \textsc{Gadget} & 100 & $2 \times 1504^{3}$ & $1.81 \times 10^{6}$ & $9.7 \times 10^{6}$ & 127 & 0 \\
        SIMBA (Flagship)       & \textsc{Gizmo}  & 100 & $2 \times 1024^{3}$ & $1.20 \times 10^{7}$ & $9.6 \times 10^{7}$ & 249 & 0 \\
        TNG100 (TNG100-1)      & \textsc{Arepo}  & 75  & $2 \times 1820^{3}$ & $1.40 \times 10^{6}$ & $7.5 \times 10^{6}$ & 127 & 0 \\
        \bottomrule
    \end{tabular}
\end{table*}

\subsection{Black Hole Feedback Models}\label{ssec:bh_feedback_models}

This section provides a concise overview of the black hole feedback models utilized in the three simulations. For comprehensive details, we direct the readers to \cite{crain2015}, \cite{McCarthy10} and \cite{McCarthy11} for EAGLE, \cite{simba} and \cite{alcazar17a} for SIMBA, and \cite{rweinberger2018} for TNG100. A convenient summary covering all three can be found in \cite{SureshBlanton2026}.

To simulate black hole growth, each of these models incorporates a variation of Bondi accretion into its subgrid physics (\citealt{hoyle_littleton_1939, bondi_hoyle1944, bondi1952}), capping the maximum growth by using the Eddington accretion rate, $\dot{M}_{\rm Edd}$. SIMBA includes an additional ``torque-limited accretion'' mode (\citealt{hopkins&quataert11, alcazar17a}) for the accretion of cold gas ($T < 10^{5}$ K). This specific mode claims to track the loss of gas angular momentum, which is essential for accurately modeling gas infall from galactic scales down to parsec scales.

The Bondi and Eddington mass accretion rates are given by:

\begin{eqnarray}\label{eq:bondi}
    \dot{M}_{\rm Bondi} &=& \epsilon_{\rm m} \frac{4 \pi G^2 M_{\rm BH}^2 \rho}{\left( v^2 + c_{\rm s}^2 \right)^{3/2}} \\
    \dot{M}_{\rm Edd} &=& \frac{4 \pi G M_{\rm BH} m_{\rm p}}{\epsilon_{\rm r} \sigma_{\rm T} c}
\end{eqnarray}

Here, G is Newton's gravitational constant, $M_{\rm BH}$ is the mass of the black hole, $\rho$ is the gas density in the vicinity of the black hole, $v$ is the speed of the black hole relative to the gas, $\epsilon_{\rm r}$ is the radiative efficiency of the black hole, $m_{\rm p}$ is the mass of the proton, $\sigma_{\rm T}$ is the Thompson cross section, $c_{\rm s}$ is the local sound speed and $c$ is the speed of light. The parameter $\epsilon_{\rm m}$ represents the efficiency of gas transport from the accretion disk to the central black hole. While this parameter is only explicitly included in the SIMBA model, we introduce it here for generalized completeness. SIMBA calibrates $\epsilon_{\rm m}$ to 0.1; for EAGLE and TNG100, we set $\epsilon_{\rm m} = 1$, additionally setting $v = 0$ for TNG100.

While the black hole mass accretion rate $\dot{M}_{\rm BH}$ relies on the Bondi accretion rate $\dot{M}_{\rm Bondi}$ across all three simulations, the exact functional dependencies vary as follows:

\begin{description}
    \item[\normalfont For EAGLE] 
    \begin{equation}
        \dot{M}_{\rm BH} = \min \left(\dot{M}_{\rm Bondi} [(c_{\rm s}/ V_{\phi})^{3}/C_{\rm visc}], \dot{M}_{\rm Bondi} \right),
    \end{equation}
    where $V_{\phi}$ is the circular speed of gas around the black hole and $C_{\rm visc}$ is a tunable gas viscosity coefficient parameter. $\dot{M}_{\rm BH}$ is allowed a maximum value of $\dot{M}_{\rm Edd}$.
    
    \item[\normalfont For SIMBA] 
    \begin{equation}
        \dot{M}_{\rm BH} = (1 - \epsilon_{\rm r})(\dot{M}_{\rm Torque} + \dot{M}_{\rm Bondi}),
    \end{equation}
    where $\dot{M}_{\rm Torque}$ is the torque limited accretion rate and has an upper limit of $3 \dot{M}_{\rm Edd}$ and $\dot{M}_{\rm Bondi}$ has an upper limit of $\dot{M}_{\rm Edd}$.
    
    \item[\normalfont For TNG100] 
    \begin{equation}
        \dot{M}_{\rm BH} = \min \left(\dot{M}_{\rm Bondi}, \dot{M}_{\rm Edd} \right)
    \end{equation}
\end{description}

Further, the black hole feedback models of the three simulations are employed as follows. EAGLE utilizes a single thermal feedback mode designed to encapsulate both the ejective effects of radiative AGN and the preventative effects of radio AGN (\citealt{McCarthy10, McCarthy11, crain2015}). In contrast, SIMBA implements two distinct feedback modes---one radiative and one radio---both of which are applied as kinetic feedback. TNG100 similarly features two modes, but applies the radiative mode as thermal feedback and the radio mode as kinetic feedback. Regardless of these specific implementations, the total black hole feedback energy or the bolometric luminosity for all three simulations is defined as:

\begin{equation}\label{eq:Lbol}
    L_{\rm Bol} = K \dot{M}_{\rm BH}c^2
\end{equation}

where,

\begin{itemize}
    \item For EAGLE: $K = \epsilon_{\rm r}$ with $\epsilon_{\rm r} = 0.1$.
    \item For SIMBA: $K = \epsilon_{\rm r} = 0.1$ for both radiative and radio mode feedback.
    \item For TNG100: $K = \epsilon_{\rm r}$ for the radiative mode and $K = \epsilon_{\rm kin}$ for the radio mode. $\epsilon_{\rm r} = 0.2$, and $\epsilon_{\rm kin}$ is a function of $\rho$, the gas density in the black hole vicinity.
\end{itemize}

In EAGLE and the radiative mode of TNG100, a factor $\epsilon_{\rm f}$ is used to define the fraction of output energy that couples back to the galaxy. Similarly, in SIMBA, only a fraction of the total energy is coupled back to the host galaxy. However, because our observations estimate the dust-corrected bolometric luminosity of the AGN (\citealt{blanton2026}), these theoretical coupling factors are not used in our computation of $L_{\rm Bol}$ in the simulations.

\subsubsection{TNG100 Black Hole Feedback Model}

Since we will look at the TNG100 black hole feedback model slightly more deeply in this study, here is a brief description of it. TNG100 includes two distinct modes of black hole feedback: a radiative mode and a radio mode. The transition between these modes depends on both the black hole accretion rate and the black hole mass.

More specifically, the transition threshold is defined by Equation~\ref{eq:default_transition_chi}:
\begin{equation}\label{eq:default_transition_chi}
\chi = \min\left(\chi_{0}\left(\frac{M_{\rm BH}}{10^{8}\,M_{\odot}}\right)^{\beta},\, 0.1\right).
\end{equation}
Here, $\chi_{0}$ and $\beta$ are free parameters with default values of $0.002$ and $2$, respectively.

Black holes with $\dot{M}_{\rm BH}/\dot{M}_{\rm Edd} > \chi$ operate in the radiative mode, whereas those with $\dot{M}_{\rm BH}/\dot{M}_{\rm Edd} < \chi$ operate in the radio mode.

In \cite{SureshBlanton2026}, we demonstrated that the default transition prescription adopted in TNG100 fails to reproduce even the qualitative observed trends in $F_{\rm AGN}$ for radio AGN. In that work, we further investigated whether varying the transition-threshold parameters, $\chi_{0}$ and $\beta$, could bring the TNG100 predictions for $F_{\rm AGN}(M_{\star}, \mathrm{sSFR})$ into agreement with the observational constraints for radio AGN. We found that introducing separate transition thresholds for star-forming and quiescent galaxies substantially improved the agreement between the TNG100 predictions and the observed $F_{\rm AGN}$ constraints. However, we also argued that such a modification would likely degrade the agreement with the observed galaxy stellar mass function, one of the primary observables used to calibrate the model in the first place. This led us to conclude that the current TNG100 framework may be unable to simultaneously reproduce both the observed radio AGN $F_{\rm AGN}$ trends and the observed galaxy stellar mass function.

In the present study, we further show that the modifications introduced in \cite{SureshBlanton2026} to improve agreement with the observed radio AGN $F_{\rm AGN}$ trends not only are likely to worsen the agreement with the observed galaxy stellar mass function, but also degrade the agreement with the observed $F_{\rm AGN}$ trends for radiative AGN (see Section~\ref{ssec:modified_tng100}).

\section{Data and Methodology}\label{sec:data_and_methods}
\subsection{Data and Methodology Overview}\label{ssec:d&m_overview}

We utilize publicly available galaxy catalogs and particle snapshot data for all three simulations: EAGLE (\citealt{mcalpine2015}), SIMBA (\citealt{simba}), and TNG100 (\citealt{illustristng2019}). These catalogs provide essential galaxy properties, such as the stellar mass $M_{\star}$ and star formation rate ($\mathrm{SFR}$), allowing us to straightforwardly compute the specific star formation rate ($\mathrm{sSFR} = \mathrm{SFR}/M_{\star}$). Our observational constraint from \cite{blanton2026} is valid within the stellar mass range of  $10 \leq \log_{10} \left( M_{\star}/ M_{\odot} \right) \leq 12$, and we impose the same stellar mass cut across all three simulations for this study.

For EAGLE, SIMBA and TNG100, we compute the Eddington ratio for the central black hole in each galaxy. Here, the Eddington ratio $\lambda$ is defined as the ratio of the bolometric luminosity $L_{\rm Bol}$ to the Eddington luminosity $L_{\rm Edd}$ of the black hole, expressed as follows:

\begin{equation}\label{eq:Ledd}
     L_{\rm Edd} = 1.26\times 10^{38}\times  M_{\rm BH}/ \rm {M_\odot} {\rm ~ergs~s}^{-1}, 
\end{equation}

\begin{equation}\label{eq:ER}
    \lambda = L_{\rm Bol}/ L_{\rm Edd},
\end{equation}

where $L_{\rm Bol}$ for the three simulations is given by Equation \ref{eq:Lbol}. 

To investigate the impact on our results due to the burstiness of $\dot{M}_{\rm BH}$ in simulations, we utilize both instantaneous and time-averaged values for $\dot{M}_{\rm BH}$ in EAGLE and SIMBA for the comparison. For EAGLE, instantaneous values are extracted from the particle snapshots (\citealt{eagle_particle_data}) and the past 100 Myr time-averaged values are taken from \cite{mcalpine2015}. For SIMBA, instantaneous values are extracted from the particle snapshots (\citealt{simba}) and the past 50 Myr time-averaged values are taken from \cite{nicole_thomas2021}. We find that our results do not change significantly whether instantaneous or time-averaged values are used for EAGLE and SIMBA. For TNG100, however, we carry out the analysis using only instantaneous values from the particle snapshots (\citealt{illustristng2019}). Based on our findings from the other two simulations, we expect these results would remain robust even if time-averaged values were used. Throughout the remainder of this analysis, all presented plots correspond to instantaneous values of $\lambda$, limited by the time resolution of the simulations. We consider this the more appropriate approach, as the observed Eddington ratios derived from narrow-line emission are constant on timescales of $\sim 10^{4}$-$10^5$ years, assuming a size of the narrow line region to be $\sim 3 ~\rm kpc$ - $30~\rm kpc$.

To evaluate the underlying physics of the black hole subgrid models across all three simulations, we compare their predictions for $F_{\rm AGN} (M_{\star}, \mathrm{sSFR})$ -- the AGN fraction as a function of $M_{\star}$ and sSFR above a critical Eddington ratio against our observations from \cite{blanton2026}. For the observational data, this critical Eddington ratio, $\lambda_{\rm c}$, is fixed at $10^{-3}$ throughout the comparison. In the simulations however, we allow the critical Eddington ratio to vary, to account for potential differences in the absolute scales of activity levels between the simulated and observed data. Specifically, we define a shifted critical Eddington ratio $\log_{10} (\lambda_{\rm c}^{\prime}) = \log_{10} (\lambda_{\rm c}) + \Delta \log_{10} (\lambda_{\rm c})$, where $\Delta \log_{10} (\lambda_{\rm c})$ takes values of 0, 1, and 2. Positive values of $\Delta \log_{10} (\lambda_{\rm c})$ indicate that the absolute level of AGN activity is higher in the simulation than in the observations. For EAGLE and SIMBA, positive values of $\Delta \log_{10} (\lambda_{\rm c})$ are more appropriate (see Figures \ref{fig:Fagn_eagle}, \ref{fig:Fagn_simba}), whereas for TNG100, negative values are more appropriate (see Figure \ref{fig:modified_tng100_transition}). This highlights the substantial differences in the absolute AGN activity levels that can arise even between different simulations, despite the fact that all of them are designed to reproduce the observed galaxy stellar mass function, among other observables.

\subsection{Data and Methodology: EAGLE}\label{ssec:d&m_eagle}

\subsubsection{Data Samples: EAGLE}\label{sssec:data_eagle}

For the EAGLE simulation, we use snapshots 27 ($z = 0.1$) and 28 ($z = 0$) from the Ref-L0100N1504 run. This sample contains a total of 7,313 galaxies within the stellar mass range
$10 \leq \log_{10}\left(M_{\star}/M_{\odot}\right) \leq 12$.
We make use of the EAGLE galaxy and halo catalogues (\citealt{mcalpine2015}) for these snapshots, which provide estimates of the stellar mass $M_{\star}$ and star formation rate (SFR), from which the specific star formation rate (sSFR) is straightforwardly derived.

For the black hole mass, $M_{\rm BH}$, and black hole accretion rate, $\dot{M}_{\rm BH}$, we use the EAGLE particle data corresponding to the same snapshots. To assess whether the bursty nature of $\dot{M}_{\rm BH}$ influences our results, we also consider accretion rates averaged over the preceding 100 Myr from \cite{McAlpine2017}. We find that using these time-averaged values does not significantly alter our results. Therefore, all plots presented in this paper are based on the instantaneous values of $\dot{M}_{\rm BH}$.

\subsubsection{Eddington Ratio Estimates: EAGLE}\label{sssec:er_eagle}

For each of the 7,313 galaxies in our EAGLE sample, we assign an Eddington ratio $\lambda$ using Equations \ref{eq:Lbol}, \ref{eq:ER} and \ref{eq:Ledd}. The EAGLE team reports that their black hole feedback model tends to exhibit radiative mode AGN behavior for $\lambda > 0.02$, and radio mode AGN behavior for $\lambda < 0.02$ (\citealt{mcalpine2015, McCarthy10, McCarthy11}). We find that applying this threshold has a negligible impact on both our results and conclusions. Consequently, the plots shown in this paper do not include this cutoff.

Using $M_{\star}$ and sSFR from the galaxy catalogs, together with the definition of $\lambda$ adopted here, we compute
$F_{\rm AGN}\left(M_{\star},~\rm sSFR\right)$
for
$\Delta \log_{10}(\lambda_{\rm c}) = 0, 1,$ and $2$.

\subsubsection{Data Sample: SIMBA}\label{sssec:data_simba}

For SIMBA, we use the galaxy catalogs corresponding to snapshots 143--151, spanning the redshift range $0 \leq z \leq 0.14$. After imposing the stellar mass selection $10 \leq \log_{10}\left(M_{\star}/M_{\odot}\right) \leq 12$, the sample contains 91,422 galaxies. From the catalogs, we use the black hole mass $M_{\rm BH}$, Eddington ratio $f_{\rm Edd}$, stellar mass $M_{\star}$, and star formation rate (SFR), from which we compute the specific star formation rate (sSFR).

\subsubsection{Eddington Ratio Estimates: SIMBA}\label{sssec:er_simba}

As noted earlier, SIMBA includes two distinct modes of black hole feedback. Nearly all galaxies host an AGN operating in the radiative mode with an associated value of $f_{\rm Edd}$, while a subset of galaxies additionally exhibit radio-mode feedback. In this work, we directly adopt the catalog-provided values of $f_{\rm Edd}$ as our Eddington ratio estimate, i.e. $\lambda = f_{\rm Edd}$, which is effectively equivalent to computing $\lambda$ using Equations~\ref{eq:Lbol}, \ref{eq:ER} and \ref{eq:Ledd}.

As in our analysis of EAGLE, we also test the effect of variability by using black hole accretion rates $\dot{M}_{\rm BH}$ averaged over the past 50~Myr, following \cite{nicole_thomas2021}, and compute the corresponding time-averaged Eddington ratio $\langle f_{\rm Edd} \rangle_{50\,\rm Myr}$. This is intended to account for the bursty nature of AGN accretion. We find that our results remain unchanged under this alternative definition. Therefore, throughout the remainder of this paper, we adopt the instantaneous values of $\dot{M}_{\rm BH}$ and $f_{\rm Edd}$.

Using the quantities defined above, we compute $F_{\rm AGN}\left(M_{\star}, \mathrm{sSFR}\right)$ at $\Delta \log_{10}(\lambda_{\rm c}) = 0, 1,$ and $2$ for comparison with observations.

\subsubsection{Data Sample: TNG100}\label{sssec:data_tng100}

For TNG100, we use snapshots 87--99, corresponding to the redshift range $0 \leq z \leq 0.15$. The galaxy catalogs provide global galaxy properties such as stellar mass $M_{\star}$ and star formation rate (SFR), from which we compute the specific star formation rate (sSFR). Applying the stellar mass selection $10 \leq \log_{10}(M_{\star}/M_{\odot}) \leq 12$ yields a sample of 70,820 galaxies.

For black hole properties, including the black hole mass $M_{\rm BH}$, accretion rate $\dot{M}_{\rm BH}$, we make use of the particle-level data at the corresponding snapshots. Given that EAGLE and SIMBA show no significant changes in the results when time-averaged accretion rates are used, we consider it reasonable to adopt the instantaneous accretion rates, $\dot{M}_{\rm BH}$, for TNG100 as well. We therefore use instantaneous values for TNG100 throughout this work.

\subsubsection{Eddington Ratio Estimates: TNG100}\label{sssec:er_tng100}

For the galaxies operating in the radiative mode of feedback, we ascribe an Eddington ratio $\lambda$ based on Equations \ref{eq:Lbol}, \ref{eq:Ledd} and \ref{eq:ER}. Using the galaxy and black hole properties defined above, we compute $F_{\rm AGN} \left(M_{\star}, sSFR \right)$ for TNG100 at $\Delta \log_{10}(\lambda_{\rm c}) = 0, 1,$ and $2$ and compare that to our observational.

\section{Results}\label{sec:results}

In this section, we present our comparison of the predicted $F_{\rm AGN} \left( M_{\star}, ~\rm sSFR\right)$ from the EAGLE, SIMBA, and TNG100 simulations with the observational constraints described in Section~\ref{intro}. As noted earlier, we claim that this comparison is a means of assessment the underlying physics of the black hole feedback subgrid models. 

Given that systematic offsets in the absolute normalization of AGN activity may exist between simulations and observations, we focus primarily on qualitative trends in $F_{\rm AGN}$. Therefore, these trends are evaluated at $\log_{10} (\lambda_{\rm c}^{\prime}) = \log_{10} (\lambda_{\rm c}) + \Delta \log_{10} (\lambda_{\rm c})$, where $\Delta \log_{10} (\lambda_{\rm c}) = 0, 1,$ and $2$.

\subsection{Metrics for Comparison}\label{ssec:metrics_for_comparison}

For the EAGLE, SIMBA, and TNG100 simulations, and for $\Delta \log_{10} (\lambda_{\rm c}) = 0, 1,$ and $2$, we adopt the following metrics to compare with observations:

\begin{itemize}
    \item To evaluate the dependence of $F_{\rm AGN}$ on $M_{\star}$ for star-forming and quiescent galaxies, we approximate the curves as straight lines and compare their slopes to those measured in the observations. In our observations, the slope of $F_{\rm AGN} \left( M_{\star}\right)$ in log-log space is 0.14 for star-forming galaxies and -1.11 for quiescent galaxies.

    \item To assess the dependence of $F_{\rm AGN}$ on sSFR, we compute the difference in $\log_{\rm 10} \left( F_{\rm AGN}\right)$ between the best fit star forming and quiescent curves at a chosen pivot point of $\log_{\rm 10} \left( M_{\star}/ M_\odot\right) = 11$. In the observations, this value is 1.27 dex, with $F_{\rm AGN}$ for star-forming galaxies being higher than that of quiescent galaxies.
    
\end{itemize}

We emphasize that these metrics provide a primarily qualitative assessment of $F_{\rm AGN} \left(M_{\star}, \rm sSFR \right)$. For instance, a model may reproduce the observed slopes and relative deviations between the star forming and quiescent curves, while remaining offset in overall normalization; such a model would still be considered successful within the scope of this evaluation.

\begin{itemize}
    
    \item EAGLE: Figure~\ref{fig:Fagn_eagle} compares the EAGLE predictions for $F_{\rm AGN}(M_{\star}, {\rm sSFR})$ with our observational constraints. The bottom-right panel corresponds to $\Delta \log_{10}(\lambda_{\rm c}) = 0$, the bottom-left panel to $\Delta \log_{10}(\lambda_{\rm c}) = 1$, and the top-left panel to $\Delta \log_{10}(\lambda_{\rm c}) = 2$. In all three cases, the critical Eddington ratio adopted for the observations is fixed at $\log_{10}(\lambda_{\rm c}) = -3$. The comparison shows that, for star-forming galaxies, EAGLE successfully reproduces the observed slope of $F_{\rm AGN}(M_{\star})$ in log--log space in all three panels. In the simulations, the slope is approximately zero, indicating an almost flat dependence on stellar mass, whereas the observed slope is $\sim 0.14$. Despite this small difference, EAGLE captures the qualitative behavior of the observations for the star-forming population.
    
    However, EAGLE fails to reproduce the strong observed decline of $F_{\rm AGN}$ with increasing $M_{\star}$ for quiescent galaxies. In the observations, the slope of this relation in log--log space is $\sim -1.11$. By contrast, EAGLE predicts at best an approximately flat relation in the top-left panel and a positive slope in the other two panels.

    Furthermore, EAGLE is also unable to correctly reproduce the dependence of $F_{\rm AGN}$ on sSFR. In the observations, the difference between the two curves at the pivot point of $\log_{\rm 10} \left( M_{\star}/ M_\odot\right) = 11$ is $\sim 1.27$. In contrast, the closest EAGLE comes to matching this separation is in the top-left panel, where the difference is $\sim 0.53$.

    \item SIMBA: Figure~\ref{fig:Fagn_simba} presents the corresponding comparison for SIMBA, with panels analogous to those in Figure~\ref{fig:Fagn_eagle}. The results indicate that SIMBA predicts a qualitatively different trend for star-forming galaxies compared to the observations. Whereas the observed slope for star-forming galaxies is positive ($\sim 0.14$), SIMBA yields a negative slope in all three panels, although the predicted decline is relatively shallow ($\sim -0.35$ in the bottom two panels).

    For quiescent galaxies, SIMBA predicts a slope in all three panels that is qualitatively consistent with the observations, although it is not as steep. The bottom-right panel provides the closest match to the data, yielding a slope of $\sim -0.59$.
    
    Finally, SIMBA performs reasonably well in reproducing the $sSFR$ dependence of $F_{\rm AGN}$. The difference in $\log_{\rm 10} \left( F_{\rm AGN}\right)$ at $\log_{\rm 10} \left( M_{\star}/ M_\odot\right) = 11$ in the observations is $\sim 1.27$. SIMBA comes closest to the observations in the top-left panel, with a value of $\sim 1.62$.
    
    \item TNG100: Figure~\ref{fig:Fagn_illustris} presents the corresponding comparison for TNG100, with panels analogous to those in Figure~\ref{fig:Fagn_eagle}. The figure shows that TNG100 performs poorly relative to the observational constraints for both star-forming and quiescent galaxies. The model predicts a strong decline of $F_{\rm AGN}$ with $M_{\star}$ for star-forming galaxies while the observed trend is nearly flat. However, it underpredicts the AGN fraction in the highest-mass bin of star-forming galaxies where the observed fraction is quite high ($\sim 30\%$). In addition, the model predicts almost no AGN in quiescent galaxies, despite the existence of observed data in this regime. 
    
    Furthermore, across all three panels at $\log_{\rm 10} \left( M_{\star}/ M_\odot\right) = 11$, TNG100 predicts a difference in $\log_{\rm 10} \left( F_{\rm AGN}\right)$ between the star forming and quiescent curves to be $> 2.22$, which is significantly higher than the observed value.
    
\end{itemize}

To summarize, EAGLE, SIMBA and TNG100 are unable to even qualitatively reproduce the observed $F_{\rm AGN} \left(M_\star, sSFR \right)$ trends for radiative AGN. EAGLE is able to correctly predict a roughly flat trend of $F_{\rm AGN}$ against $M_\star$ for star forming galaxies, but is unable to predict the declining trend of $F_{\rm AGN}$ with $M_{\star}$ for quiescent galaxies. SIMBA on the other hand is able to reproduce a qualitatively similar trend to the observations for quiescent galaxies, but the slope is not sufficiently steep. Also, it predicts a qualitatively incorrect trend for star-forming galaxies. In contrast, TNG100 performs poorly relative to the observations for both star-forming and quiescent populations. In particular, it is unable to predict a flat $F_{\rm AGN} \left(M_\star \right)$ trend for star-forming galaxies and it predicts almost no AGN in quiescent galaxies, despite the clear presence of a significant number of detected AGN in the observational data over the same mass range.

\subsection{Modified TNG100 Model}\label{ssec:modified_tng100}

In this section, we investigate whether the TNG100 feedback model can be modified to force the prediction of the radiative AGN fraction, $F_{\mathrm{AGN}}\left( M_{\star}, \mathrm{sSFR} \right)$, into alignment with observations. As discussed in Section~\ref{sec:sims_and_bhfeedback}, TNG100 utilizes a black hole mass dependent Eddington-scaled accretion rate to govern the transition between the radio and radiative feedback modes. In Figure~\ref{fig:modified_tng100_transition}, the transition threshold for the default TNG100 model is indicated by the grey dashed line; black holes above this line operate in the radiative mode, while those below it operate in the radio mode. Here, we examine whether this threshold can be altered by tuning the parameters in Equation~\ref{eq:default_transition_chi} to bring the TNG100 $F_{\mathrm{AGN}}$ predictions into closer agreement with observations. Specifically, we modify the model parameterization as:

\begin{equation}\label{eq:modified_transition_chi}
\chi_{\mathrm{Q}}, \chi_{\mathrm{SF}} = \min\left( \chi_{0} \left( \frac{M_{\mathrm{BH}}}{M_{0}} \right)^{\beta}, \chi_{\mathrm{max}} \right)
\end{equation}

and vary the parameter set 
\[
\theta = \{ \chi_0, M_0, \beta, \chi_{\mathrm{max}}, \Delta \log_{10} (\lambda_{\mathrm{c}}) \}
\]
separately for star-forming and quiescent galaxies. Our objective is to minimize the $\chi^{2}$ error between the predicted and observed $F_{\mathrm{AGN}} \left( M_{\star} \right)$ curves, treating both galaxy populations independently. Note that $\Delta \log_{10} \left( \lambda_{\mathrm{c}} \right)$ is not a part of the default model, but is included here for the process of optimization.

Figure~\ref{fig:modified_tng100_transition} also displays the modified transition lines for radiative AGN, determined separately for the star-forming and quiescent populations. These modified models correspond to the following best-fit parameter sets:
\begin{itemize}
    \item \textbf{Star-forming galaxies:} $\theta_{\mathrm{SF}} = \{ \chi_{0}=0.09,\, M_{0} = 10^{10.48},\, \beta = 1.95,\, \chi_{\mathrm{max}} = 0.0008,\, \Delta \log_{10} (\lambda_{\mathrm{c}}) = -0.89 \}$
    \item \textbf{Quiescent galaxies:} $\theta_{\mathrm{Q}} = \{ \chi_{0}=0.08,\, M_{0} = 10^{10.12},\, \beta = 2.00,\, \chi_{\mathrm{max}} = 0.001,\, \Delta \log_{10} (\lambda_{\mathrm{c}}) = -0.93 \}$
\end{itemize}

Figure~\ref{fig:modified_tng100_predictions} shows the resulting $F_{\mathrm{AGN}} \left(M_{\star}, \mathrm{sSFR} \right)$ predictions from the modified model plotted against the observations. Compared to the default TNG100 model, the agreement with observational data is substantially improved. Crucially, the modified model now predicts a non-zero AGN fraction in quiescent galaxies, and, except for the lowest and highest $M_\star$ bins, successfully reproduces the observed decline of $F_{\mathrm{AGN}}$ with increasing $M_{\star}$.

Nevertheless, Figure~\ref{fig:modified_tng100_predictions} also reveals that even with the optimized parameter set $\theta_{\mathrm{SF}}$, the model fails to reproduce the flat trend of $F_{\mathrm{AGN}}$ against $M_{\star}$ observed in star-forming galaxies. This mismatch is an artifact of the sharp drop-off in the underlying distribution of star-forming galaxies caused by the default transition line at $\log_{10} \left(M_{\mathrm{BH}}/ M_\odot \right) \sim 8.5$.

We performed a similar analysis in \citet{SureshBlanton2026} for the radio AGN population in TNG100 and found that it was possible to elicit better agreement with the radio AGN $F_{\rm AGN}$ constraint (see Figure~9 of \citealt{SureshBlanton2026}). However, we also noted that within a self--consistent simulations, tuning the model to match radio AGN trends might inherently degrade its agreement with the observed galaxy stellar mass function---a key observable that the model was calibrated to in the first place. The fundamental caveat raised in \citet{SureshBlanton2026} still applies here: forcing the model to match the observed $F_{\mathrm{AGN}}$ constraints for radiative AGN will likely degrade its agreement with the galaxy stellar mass function.

Furthermore, the optimized transition threshold for radiative AGN in star-forming galaxies (Figure~\ref{fig:modified_tng100_transition}) differs substantially from that of the radio AGN population (see Figure~9 of \citealt{SureshBlanton2026}); notably, these two optimized thresholds shift away from the default TNG100 model in opposite directions. This disparity implies that forcing an improvement in the radio AGN population may inherently also worsen the performance for radiative AGN, and vice versa. Consequently, the TNG framework may be fundamentally limited in its ability to simultaneously reproduce the observed galaxy stellar mass function, and the observed radio AGN fraction trends and the observed radiative AGN fraction trends.

\subsection{Combined Radiative AGN and Radio AGN Analysis for EAGLE}\label{ssec:combined_analysis_eagle}

\subsubsection{Combined $F_{\rm AGN} \left(M_\star, sSFR \right)$ Observational Constraint}

As stated in Section~\ref{ssec:bh_feedback_models} and described by \citet{crain2015} and \citet{schaye2015}, EAGLE does not explicitly differentiate between radiative- and radio-mode feedback. Instead, the simulation employs a single mode of thermal feedback that mimics the functionality of both modes across their appropriate accretion regimes: displaying radiative-mode behavior at Eddington-scaled accretion rates of $\sim 1\%$, and displaying radio-mode behavior at lower accretion rates (\citealt{crain2015, McCarthy10, McCarthy11}). Because EAGLE does not distinctively separate these two modes, we construct a combined observational baseline to ensure a fair comparison with the simulation's predictions.

EAGLE outputs a bolometric luminosity given by Equation~\ref{eq:Lbol} and assumes that a constant fraction, $\epsilon_{\mathrm{f}} = 0.15$, of this power couples to the host galaxy across the entire black hole and galaxy population. While this assumption of a constant coupling efficiency may oversimplify real physical systems, it is the explicit prescription adopted within the EAGLE framework, and we apply it to our observations, as described below.

In the narrow-line observational constraints for radiative AGN \citep{blanton2026}, the Eddington ratio $\lambda$ is defined as $\lambda = L_{\mathrm{Bol}}/ L_{\mathrm{Edd}}$, where $L_{\mathrm{Bol}}$ is the bolometric luminosity representing the total radiative power emitted by the AGN. Conversely, for the radio AGN constraints \citep{SureshBlanton2024}, the Eddington ratio is defined as $\lambda = P_{\mathrm{cav}}/L_{\mathrm{Edd}}$, where $P_{\mathrm{cav}}$ is the average kinetic jet power required to inflate cavities in the X-ray emitting hot gas surrounding the galaxy (\citealt{dun&fabian2006, fabian12, SureshBlanton2024}). In other words, $P_{\mathrm{cav}}$ reflects the AGN power that actually couples to the host galaxy environment. 

To merge the $F_{\mathrm{AGN}} \left( M_{\star}, \mathrm{sSFR}\right)$ constraints from these two distinct populations into a unified quantity, the definition of $\lambda$ must be treated consistently regarding whether it measures total emitted power or coupled power. To achieve this, we re-derive the radiative AGN observational constraint shown in Figure~\ref{fig:Fagn_constraint} by scaling its Eddington ratios by the EAGLE coupling efficiency, $\epsilon_{\mathrm{f}} = 0.15$. We then sum this rescaled radiative constraint with the radio AGN constraint from \citet{SureshBlanton2026} to produce the combined observational constraint. The resulting combined $F_{\mathrm{AGN}}$ constraint for a threshold of $\lambda > \lambda_{\mathrm{c}} = 10^{-3}$ is plotted across all three panels of Figure~\ref{fig:combined_Fagn_eagle} for direct comparison against EAGLE's predictions. Table \ref{tab:combined_Fagn_constraint} provides the values for the combined observational constraint.

\begin{table*}[t]
    \centering
    \begin{tabular}{ccccc}
        \toprule
        \multirow{2}{*}{\textbf{$\log_{\rm 10} \left( M_\star/ M_\odot\right)$}} & \multicolumn{2}{c}{$\log F_{\text{AGN}}^{\text{SF}}$} & \multicolumn{2}{c}{$\log F_{\text{AGN}}^{\text{Q}}$} \\ 
        \cmidrule(lr){2-3} \cmidrule(lr){4-5}
                                           & Lower Bound & Upper Bound & Lower Bound & Upper Bound \\ 
        \midrule
        10 -- 10.4        & $-1.24$     & $-0.11$     & $-1.91$     & $-0.41$     \\ 
        10.4 -- 10.8      & $-1.34$     & $-0.56$     & $-2.43$     & $-1.41$     \\ 
        10.8 -- 11.2      & $-1.18$     & $-0.75$     & $-1.75$     & $-1.40$     \\ 
        11.2 -- 11.6      & $-0.99$     & $-0.70$     & $-1.45$     & $-0.88$     \\ 
        11.6 -- 12        & $-0.65$     & $-0.15$     & $-1.44$     & $-0.24$     \\ 
        \bottomrule
    \end{tabular}
    \caption{1$\sigma$ bounds of $\log_{\rm 10} \left(F_{\rm AGN} \right)$ in bins of $\log_{\rm 10} \left( M_\star/ M_\odot\right)$
    for the combined radio AGN and radiative AGN observational constraint, for star forming and quiescent galaxies.}
    \label{tab:combined_Fagn_constraint}
\end{table*}

\subsubsection{Combined $F_{\rm AGN} \left(M_\star, sSFR \right)$ Comparison Results}

Figure~\ref{fig:combined_Fagn_eagle} also displays the EAGLE prediction for $F_{\mathrm{AGN}} \left(M_{\star}, \mathrm{sSFR} \right)$ at $\Delta \log_{\rm 10} \left( \lambda_{\rm c}\right) = $ 0, 1 and 2. Here, the Eddington ratio $\lambda$ represents the ratio of the AGN power coupled to the host galaxy relative to the Eddington luminosity of the black hole:

\begin{equation}
    \lambda = \frac{\epsilon_{\mathrm{f}} \epsilon_{\mathrm{r}}\dot{M}_{\mathrm{BH}}c^{2}} {L_{\mathrm{Edd}}}
\end{equation}

\begin{figure*}[h]
    \centering
    \includegraphics[width = 7.2in]{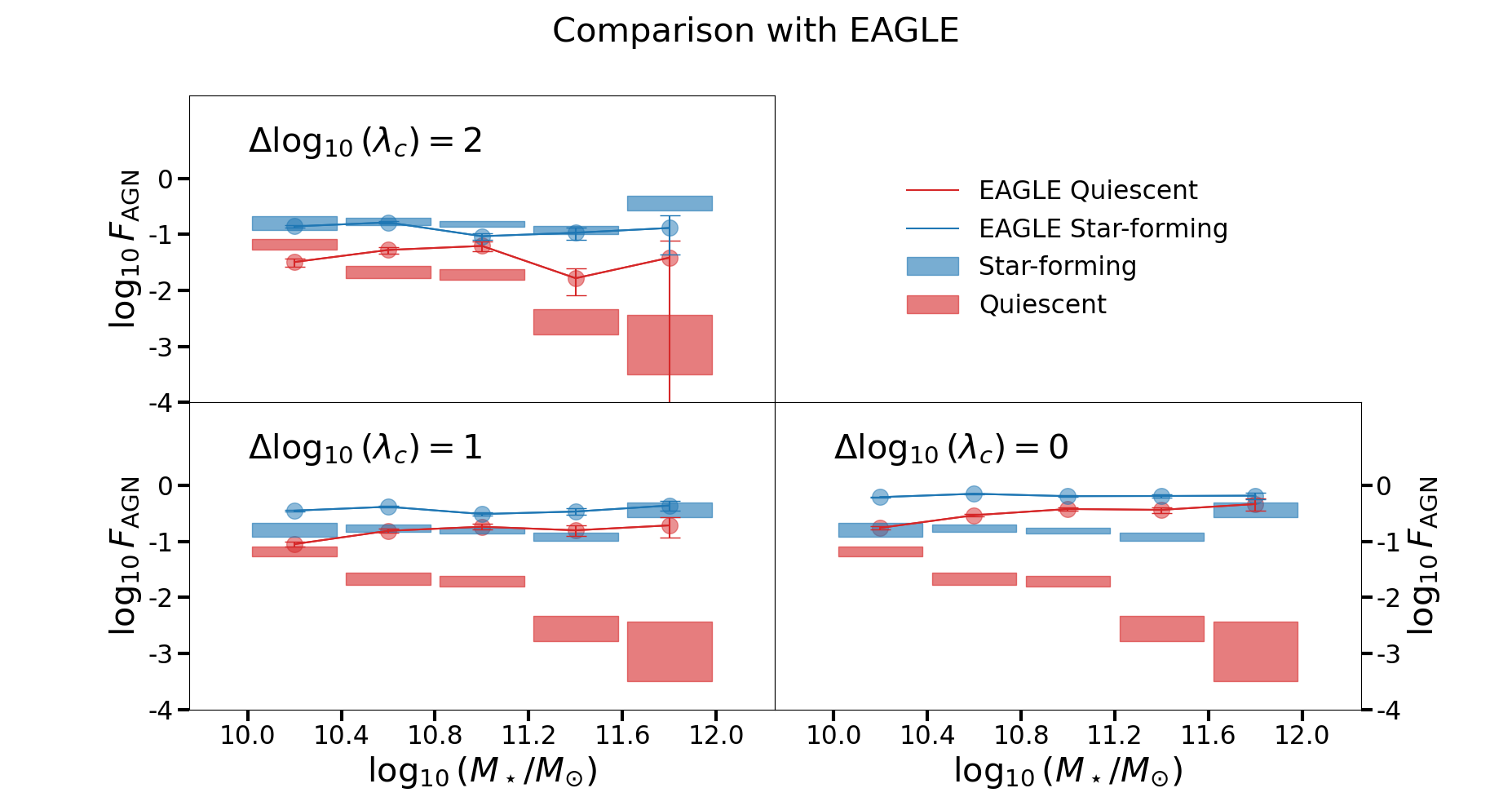}
    \caption{Comparison of the EAGLE simulation's predictions for \( F_{\rm AGN}(M_\star) \), including \( 1\sigma \) binomial uncertainties for both star-forming and quiescent galaxy populations, against the observational constraints shown in Figure~\ref{fig:Fagn_constraint}. Red points represent quiescent galaxies in EAGLE and blue points represent star-forming galaxies in EAGLE, separated at $\log_{10}(\mathrm{sSFR}/\mathrm{yr}^{-1}) = -11$. To account for potential differences in the normalization of $\lambda$ between observations and simulations, we present three subplots corresponding to different values of $\Delta \log_{10}(\lambda_{\rm c})$.}
    \label{fig:Fagn_eagle}
\end{figure*}

Note that the coupling efficiency $\epsilon_{\mathrm{f}} = 0.15$ is explicitly included in this definition of $\lambda$, whereas it was omitted in Equation~\ref{eq:Lbol}. Consequently, the EAGLE $F_{\mathrm{AGN}} \left(M_{\star}, \mathrm{sSFR} \right)$ curves in Figure~\ref{fig:combined_Fagn_eagle} correspond to the same curves shown in Figure~\ref{fig:Fagn_eagle}, but with the original Eddington ratios scaled downward by a factor of 0.15. 

As seen in the figure, EAGLE's predictions for the combined radiative and radio AGN fraction trends are in reasonable agreement with the observed data at $\Delta \log_{10} \left( \lambda_{\mathrm{c}} \right) = 0$, but perform poorly at $\Delta \log_{10} \left( \lambda_{\mathrm{c}} \right) = 2$. Crucially however, at $\Delta \log_{10} \left( \lambda_{\mathrm{c}} \right) = 1$ (shown in the bottom-left panel), the combined EAGLE predictions become almost perfectly consistent with our observations, with a minor deviations appearing only in the central $M_{\star}$ bins for quiescent galaxies. This agreement might indicate that the total AGN feedback predicted by EAGLE is consistent with our observational constraints, given EAGLE's choice of a constant coupling factor ($\epsilon_{\mathrm{f}} = 0.15$) and our adopted scaling choices---namely, the bolometric corrections for narrow-line AGN (see \citealt{blanton2026, netzer2019}) and the radio luminosity to jet power scaling (\citealt{SureshBlanton2024, cavagnolo2010}). In other words, this demonstrates that EAGLE successfully passes our independent test of its subgrid AGN feedback prescription.

\begin{figure*}[h]
    \centering
    \includegraphics[width = 7.2in]{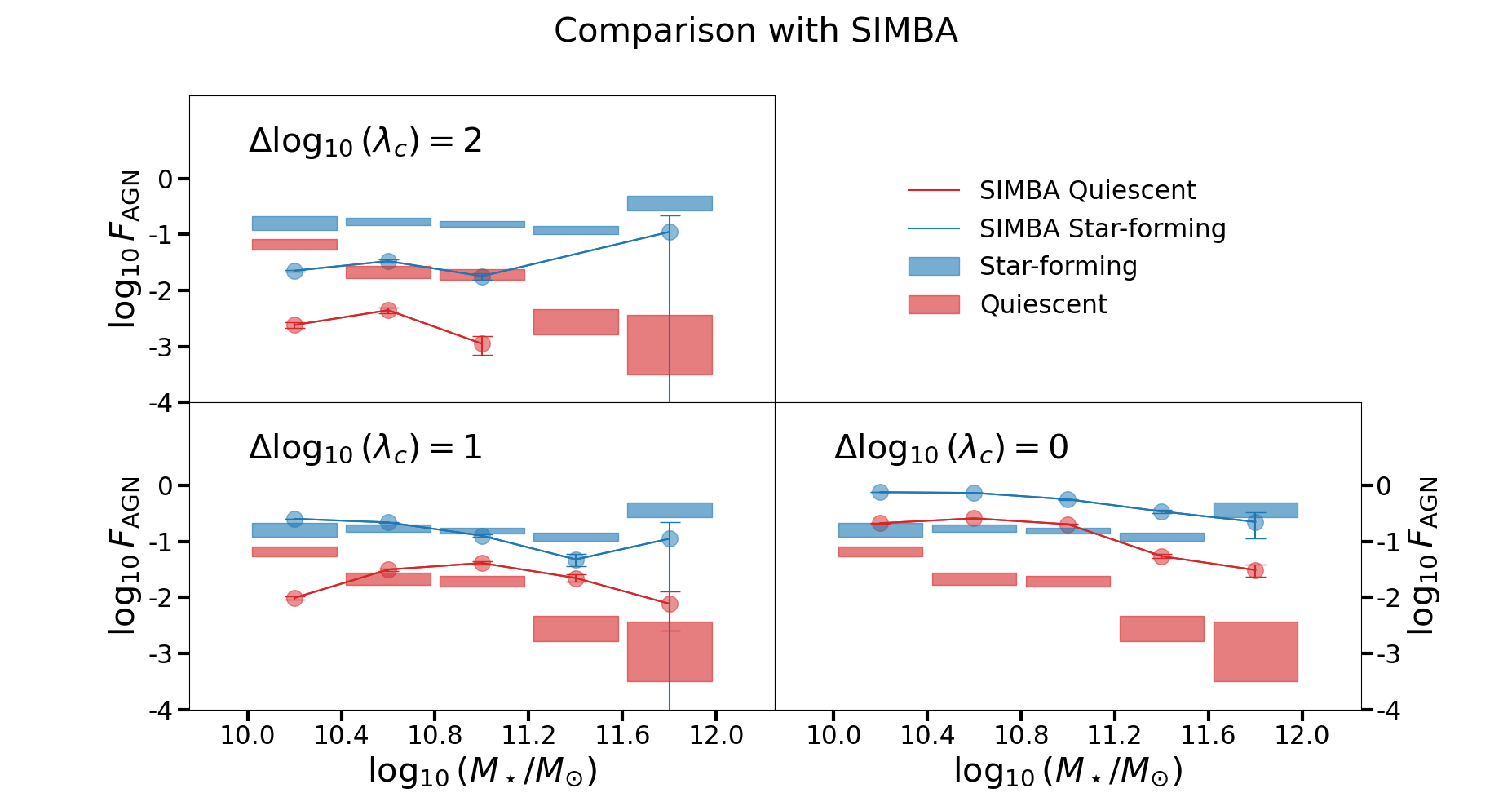}
    \caption{Similar to Figure \ref{fig:Fagn_eagle}, for SIMBA. The dividing line between quiescent and star-forming galaxies in SIMBA is set at \( \log_{10} \left( \mathrm{sSFR}/\mathrm{yr}^{-1} \right) = -11.5 \), consistent with SIMBA's internal classification.}
    \label{fig:Fagn_simba}
\end{figure*}

\begin{figure*}[h]
    \centering
    \includegraphics[width = 7.2in]{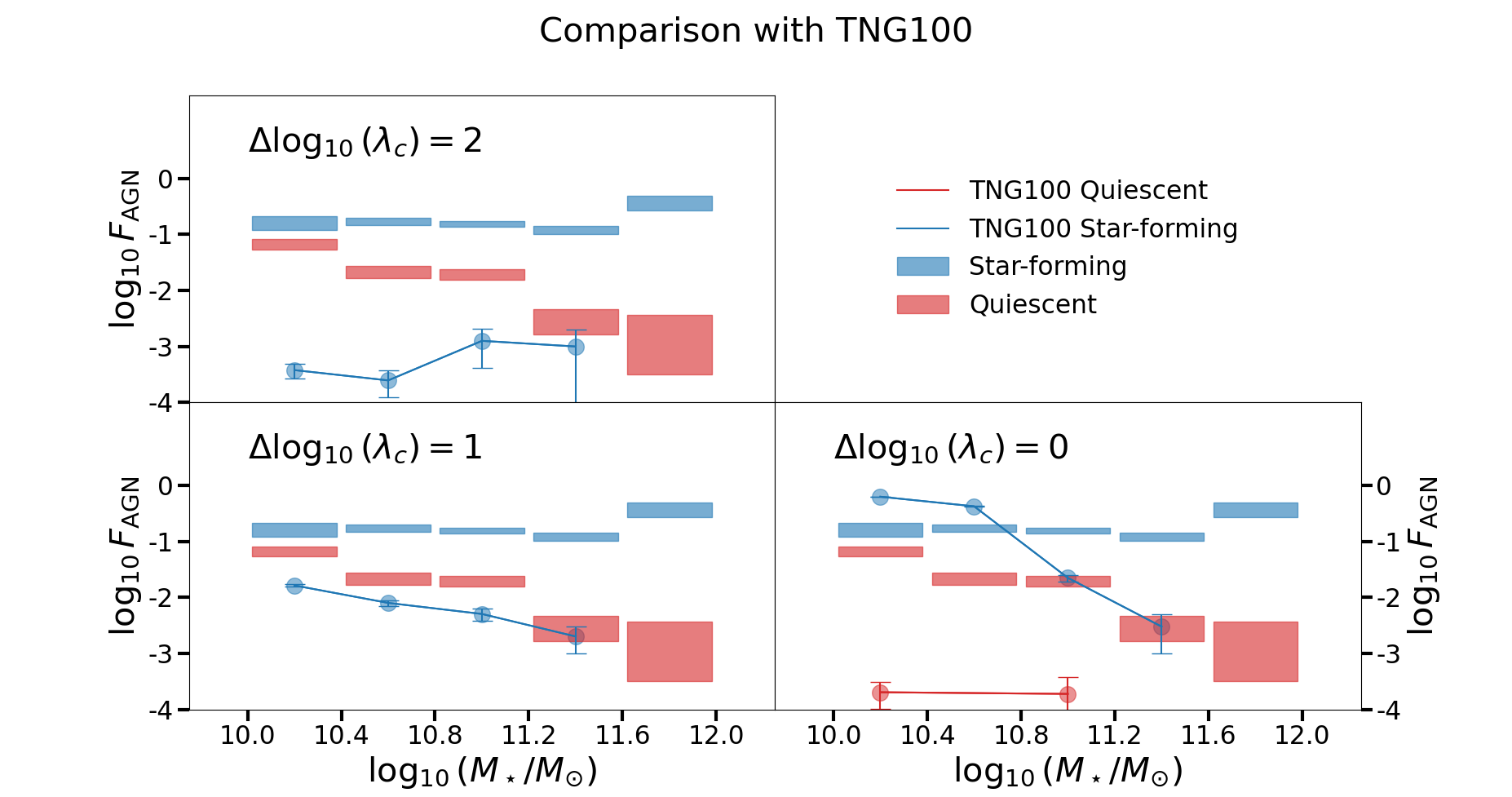}
    \caption{Similar to Figure \ref{fig:Fagn_eagle}, for TNG100. The dividing line between quiescent and star-forming galaxies in TNG100 is set at \( \log_{10} \left( \mathrm{sSFR}/\mathrm{yr}^{-1} \right) = -11.5 \), consistent with TNG100's internal classification.}
    \label{fig:Fagn_illustris}
\end{figure*}

\begin{figure*}[htbp]
  \centering
  \includegraphics[width=0.95\textwidth]{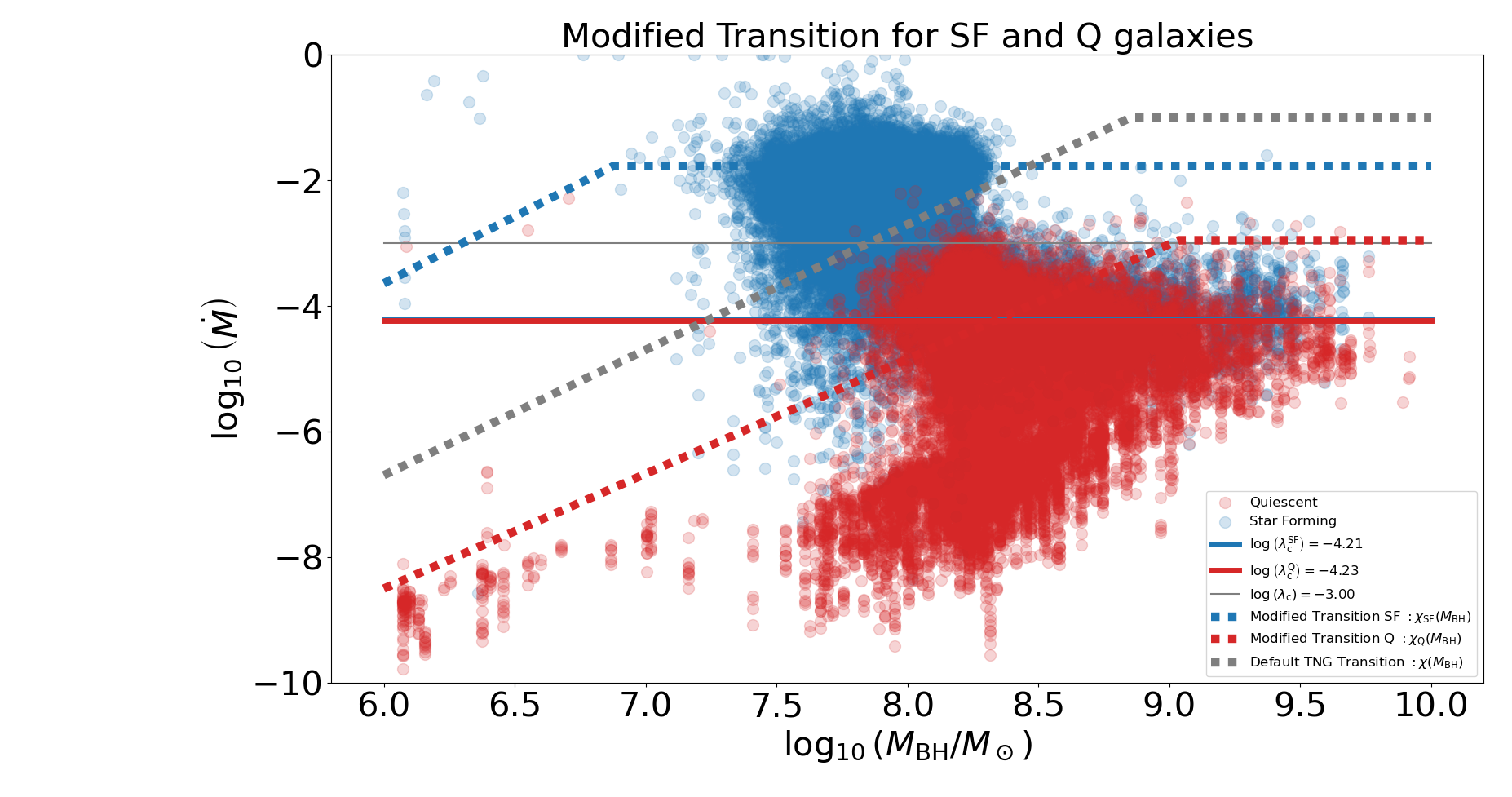}
  \caption{$\dot{M}$ versus $M_{\rm BH}$ for our TNG100 sample. The thick gray dashed line represents the default TNG100 transition curve, $\chi(M_{\rm BH})$, while the thin gray solid line corresponds to $\log_{10}(\lambda_{\rm c}) = -3$. The thick red dashed line indicates the modified transition curve for quiescent galaxies, $\chi_{\rm Q}(M_{\rm BH})$, with the associated thin red solid line representing $\log_{10}(\lambda^{\rm Q}_{\rm c}) = -4.23$. Similarly, the thick blue dashed line shows the modified transition curve for star-forming galaxies, $\chi_{\rm SF}(M_{\rm BH})$, and the thin blue solid line corresponds to $\log_{10}(\lambda^{\rm SF}_{\rm c}) = -4.21$. For a given model, galaxies above the transition line are in the radiative mode of AGN feedback, while those below the line are in the radio mode. Galaxies located above the default transition line but below the modified would host radiative AGN under the default model, but not under the modified prescriptions.}
  \label{fig:modified_tng100_transition}
\end{figure*}

\begin{figure*}[htbp]
  \centering
  \includegraphics[width=0.95\textwidth]{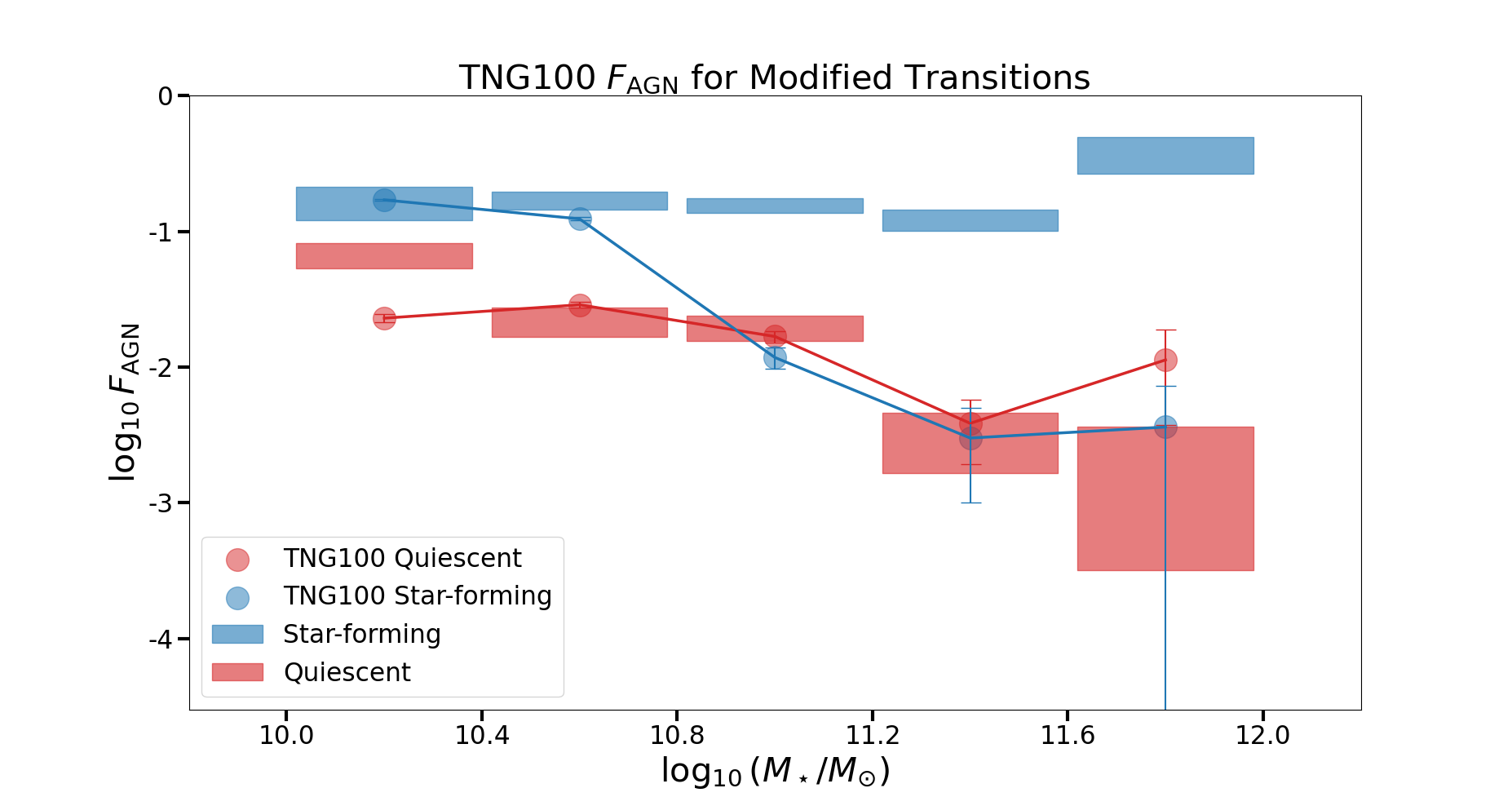}
  \caption{Predictions for $F_{\rm AGN}(M_\star)$ under the modified models for quiescent and star-forming galaxies shown in Figure \ref{fig:modified_tng100_transition}, compared to our observational constraint.}
  \label{fig:modified_tng100_predictions}
\end{figure*}

\begin{figure*}[h]
    \centering
    \includegraphics[width = 7.2in]{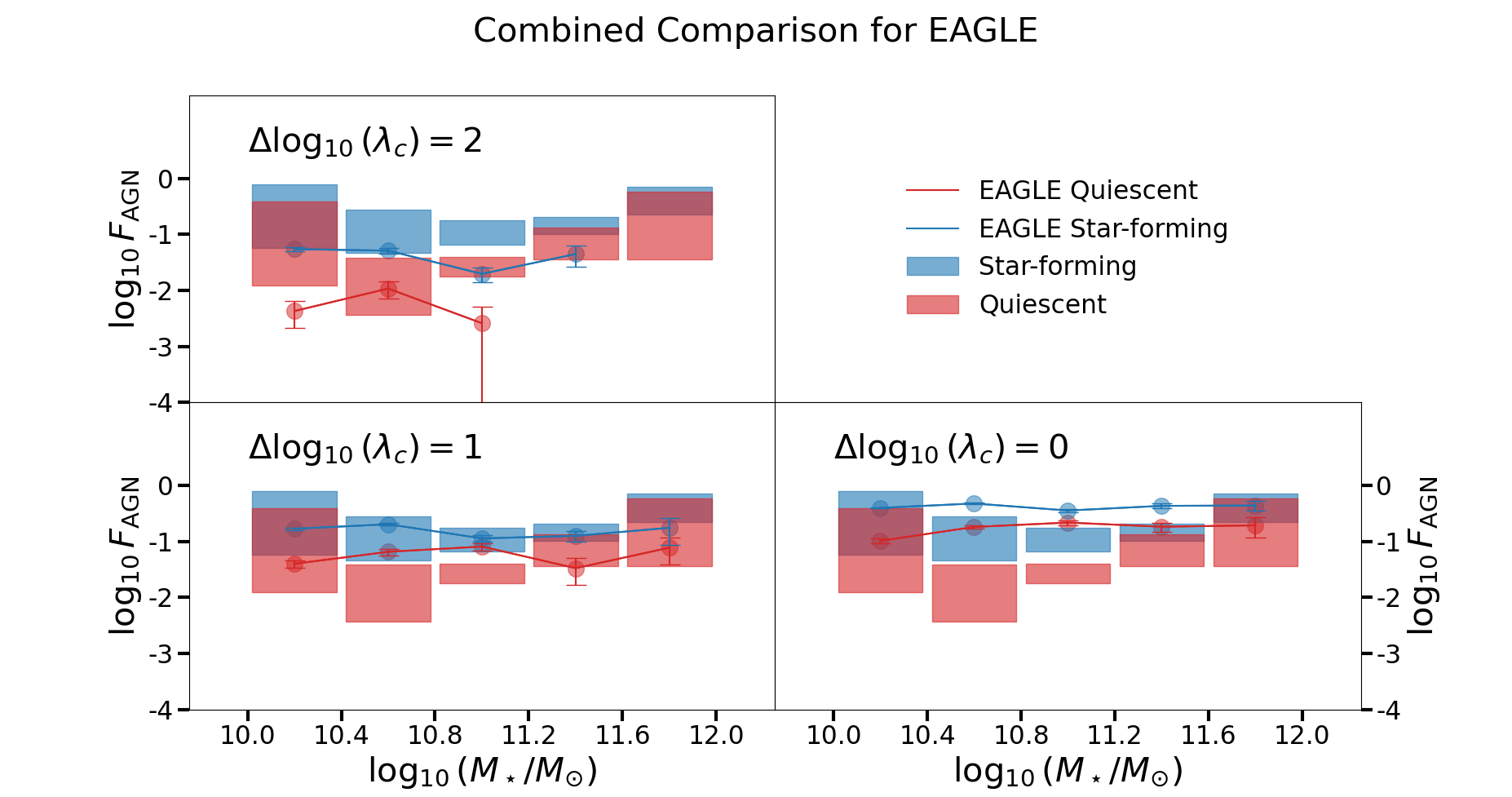}
    \caption{Comparison of the EAGLE simulation's predictions for the combined radio and radiative \( F_{\rm AGN}(M_\star, sSFR) \) including \( 1\sigma \) binomial uncertainties with the observational constraints for the same. Red represents quiescent galaxies and blue represents star-forming galaxies. To account for potential differences in the normalization of $\lambda$ between observations and simulations, we present three subplots corresponding to different values of $\Delta \log_{10}(\lambda_{\rm c})$. It can be seen that EAGLE's predictions are broadly consistent with our observations, in the bottom-left panel at $\Delta \log_{10}(\lambda_{\rm c}) = 1$.}
    \label{fig:combined_Fagn_eagle}
\end{figure*}

\section{Conclusion}\label{sec:conclusion}

AGN feedback is a field of central importance in astrophysics and cosmology, with far-reaching implications for our understanding of galaxy evolution and the Universe as a whole. However, owing to our incomplete understanding of the underlying physics, AGN feedback must be implemented through subgrid prescriptions whose physical validity remains uncertain. Despite the widespread use of these models, relatively few studies have directly examined whether they accurately predict trends in AGN activity as a function of host galaxy properties. \cite{houda2022} and \cite{habouzit2022} are studies that have carried out similar studies, but the observational constraints they use have not all been accounted for the relevant selection effects.

 In this work, we address this gap by testing the radiative mode AGN feedback prescriptions implemented in the EAGLE, SIMBA, and TNG100 simulations against our $F_{\rm AGN}(M_{\star}, {\rm sSFR})$ observational constraint (see Section \ref{intro}). $F_{\rm AGN}$ is the intrinsic fraction of galaxies hosting a radiative mode AGN with $\lambda > \lambda_{\rm c} = 10^{-3}$, where $\lambda$ is the Eddington ratio. In the observations, $F_{\rm AGN}$ drops strongly with $M_{\star}$ for quiescent galaxies while staying rough constant with $M_\star$ for star forming galaxies. Since none of these simulations were explicitly calibrated to reproduce these observed trends, this comparison constitutes an independent test of their AGN feedback prescriptions.
 
\citet{aird2019} and \citet{birchall2023} make similar measurements in of $F_{\rm AGN}$ their studies, but using AGN X-ray emission instead of narrow line emission. Their measurements of the dependence of $F_{\rm AGN}$ on $M_{\star}$ qualitatively agree with ours for star-forming galaxies, although there are quantitative differences. For quiescent galaxies, however, their measurements qualitatively disagree with ours, as they do not observe the decline that we find. The origin of this discrepancy is unclear and warrants further investigation. Nevertheless, in this study, we adopt our narrow-line observational constraint as the benchmark against which to test the simulations.

 In this study, we find the following:

\begin{itemize}
    \item EAGLE, SIMBA, and TNG100 are unable to reproduce the observed trends in $F_{\rm AGN}(M_{\star}, {\rm sSFR})$ for radiative AGN. Observationally, we find that $\log_{10}(F_{\rm AGN})$ remains approximately constant over the stellar mass range $10 \leq \log_{10}(M_{\star}/M_{\odot}) \leq 12$ for star-forming galaxies, while it declines sharply over the same mass range for quiescent galaxies (Figure~\ref{fig:Fagn_constraint}). None of the three simulations reproduces these trends for any choice of $\Delta \log_{10}(\lambda_{\rm c})$, a parameter introduced to account for possible differences in the absolute normalization of AGN activity between the simulations and the observations.

     Among the three models, EAGLE comes closest to reproducing the observations. It predicts an approximately flat relation between $\log_{10}(F_{\rm AGN})$ and $\log_{10}(M_{\star}/M_{\odot})$ for star-forming galaxies, in agreement with the data, but fails to reproduce the observed decline in $F_{\rm AGN}$ with stellar mass for quiescent galaxies (Figure~\ref{fig:Fagn_eagle}). SIMBA does not reproduce the observed constant $F_{\rm AGN}$--$M_{\star}$ relation; however, for quiescent galaxies it does predict a declining trend, albeit one that is substantially weaker than observed (Figure~\ref{fig:Fagn_simba}). TNG100 shows poorer agreement with the observations than both EAGLE and SIMBA: at $\lambda_{\rm c}=10^{-3}$ it predicts virtually no radiative AGN in quiescent galaxies, despite their clear detection in the observations, and it also fails to reproduce the observed flat dependence of $F_{\rm AGN}$ on $M_{\star}$ for star-forming galaxies (Figure~\ref{fig:Fagn_illustris}).

    \item For TNG100, we explore modifications to the default feedback model to assess whether improved agreement with the observed $F_{\rm AGN}$ constraints can be achieved (Section~\ref{ssec:modified_tng100}). In particular, we vary the Eddington-scaled accretion rate threshold that determines the transition between the radiative and radio modes of black hole feedback. We find that this modification leads to a modest improvement in the quiescent galaxy population, enabling the model to reproduce the observed decline in $F_{\rm AGN}$ with $M_{\star}$. However, for no choice of transition parameters does TNG100 simultaneously recover realistic predictions of $F_{\rm AGN}(M_{\star})$ for star-forming galaxies (Figures~\ref{fig:modified_tng100_transition} and \ref{fig:modified_tng100_predictions}).

    Furthermore, we find that improving agreement with the $F_{\rm AGN}$ constraints for radiative AGN can degrade agreement with the corresponding constraints for radio AGN. More generally, tuning the model to better match either the radiative or radio $F_{\rm AGN}$ constraints may worsen agreement with the observed galaxy stellar mass function, a key observable used in the original calibration of the simulation. Robustly testing these trade-offs would require rerunning the simulation with the modified feedback prescriptions, which is beyond the scope of this work.

     \item For EAGLE, which does not explicitly separate radiative and radio AGN feedback modes, we also assess its predictions by comparing to an observational constraint on total AGN feedback. For this, we sum the radiative and radio contributions to the observed $F_{\rm AGN}$ constraint and compare the resulting relation to EAGLE's predictions (Figure~\ref{fig:combined_Fagn_eagle}).

    For consistency between the two observational components, we rescale the measured Eddington ratios of the radiative AGN from \cite{blanton2026} by the EAGLE feedback coupling efficiency, $\epsilon_{\rm f}=0.15$, since the radio-mode constraints correspond to a ``coupled $\lambda$''. We apply this rescaling prior to combining the radiative and radio $F_{\rm AGN}$ measurements.

    This procedure yields an observed $\log_{10}(F_{\rm AGN})$ that is approximately constant with $\log_{10}(M_{\star}/M_{\odot})$ for both star-forming and quiescent galaxies. EAGLE predictions are consistent with the combined observational constraint at $\Delta \log_{10}(\lambda_{\rm c}) = 1$. In other words, when EAGLE's total AGN feedback power is effectively scaled down by a factor of ten, EAGLE reproduces the observed dependence of $F_{\rm AGN}$ on $M_{\star}$ and sSFR, given the adopted coupling efficiency $\epsilon_{\rm f}=0.15$, the narrow-line-to-bolometric correction for radiative AGN (see \citealt{netzer2019, blanton2026}), and the radio-to-jet power scaling for radio AGN (see \citealt{cavagnolo2010, SureshBlanton2024}). Under these assumptions, EAGLE successfully passes our independent test of its AGN feedback subgrid model.

    Although EAGLE satisfies this observational test, by design it does not capture the detailed underlying physics of the AGN feedback mechanism. Our results suggest that AGN feedback processes in the local Universe may indeed yield EAGLE-like behavior overall. However, even if this is the case, fully understanding the involved small-scale processes will require higher-resolution predictions to elucidate them further.
\end{itemize}

\section*{Acknowledgements}

We thank the EAGLE, SIMBA, and IllustrisTNG teams for making their data public. We convey our heartfelt gratitude to Nicole Thomas of University of Western Australia and Stuart McAlpine of University of Helsinki for their eagerness to share and help with time-averaged black hole accretion rate data. 

Finally, we would like to acknowledge the use of generative AI tools in the preparation of this work; specifically, ChatGPT was used to refine the prose of the manuscript for clarity, and Gemini was utilized during the development and verification of the analysis code.

EAGLE used the DiRAC Data Centric system at Durham University, operated by the Institute for Computational Cosmology on behalf of the STFC DiRAC HPC Facility (\url{www.dirac.ac.uk}); this equipment was funded by BIS National E-infrastructure capital grant ST/K00042X/1, STFC capital grant ST/H008519/1, STFC DiRAC Operations grant ST/K003267/1 and Durham University. DiRAC is part of the National E-Infrastructure. The study was sponsored by the Dutch National Computing Facilities Foundation (NCF) for the use of supercomputer facilities, with financial support from the Netherlands Organisation for Scientific Research (NWO), and the European Research Council under the European Union’s Seventh Framework Programme (FP7/2007–2013) / ERC Grant agreements 278594 GasAroundGalaxies, GA 267291 Cosmiway, and 321334 dustygal. Support was also received via the Interuniversity Attraction Poles Programme initiated by the Belgian Science Policy Office ([AP P7/08 CHARM]), the National Science Foundation under Grant No. NSF PHY11-25915, and the UK Science and Technology Facilities Council (grant numbers ST/F001166/1 and ST/I000976/1) via rolling and consolidating grants awarded to the ICC.

The SIMBA simulation was run on the DiRAC@Durham facility managed by the Institute for Computational Cosmology on behalf of the STFC DiRAC HPC Facility. The equipment was funded by BEIS capital funding via STFC capital grant nos ST/P002293/1, ST/R002371/1, and ST/S002502/1, Durham University, and STFC operations grant no. ST/R000832/1. DiRAC is part of the National e-Infrastructure.

The IllustrisTNG simulations were undertaken with compute time awarded by the Gauss Centre for Supercomputing (GCS) under GCS Large-Scale Projects GCS-ILLU and GCS-DWAR on the GCS share of the supercomputer Hazel Hen at the High Performance Computing Center Stuttgart (HLRS), as well as on the machines of the Max Planck Computing and Data Facility (MPCDF) in Garching, Germany.

Funding for the Sloan Digital Sky Survey IV has been provided by the Alfred P. Sloan Foundation, the U.S. Department of Energy Office of Science, and the Participating Institutions.

SDSS-IV acknowledges support and resources from the Center for High Performance Computing at the University of Utah. The SDSS website is \url{www.sdss4.org}.

SDSS-IV is managed by the Astrophysical Research Consortium for the Participating Institutions of the SDSS Collaboration including the Brazilian Participation Group, the Carnegie Institution for Science, Carnegie Mellon University, Center for Astrophysics | Harvard \& Smithsonian, the Chilean Participation Group, the French Participation Group, Instituto de Astrof\'isica de Canarias, The Johns Hopkins University, Kavli Institute for the Physics and Mathematics of the Universe (IPMU) / University of Tokyo, the Korean Participation Group, Lawrence Berkeley National Laboratory, Leibniz Institut f\"ur Astrophysik Potsdam (AIP), Max-Planck-Institut f\"ur Astronomie (MPIA Heidelberg), Max-Planck-Institut f\"ur Astrophysik (MPA Garching), Max-Planck-Institut f\"ur Extraterrestrische Physik (MPE), National Astronomical Observatories of China, New Mexico State University, New York University, University of Notre Dame, Observat\'orio Nacional / MCTI, The Ohio State University, Pennsylvania State University, Shanghai Astronomical Observatory, United Kingdom Participation Group, Universidad Nacional Aut\'onoma de M\'exico, University of Arizona, University of Colorado Boulder, University of Oxford, University of Portsmouth, University of Utah, University of Virginia, University of Washington, University of Wisconsin, Vanderbilt University, and Yale University.

\bibliography{refs.bib}

\end{document}